# On regression analysis with Padé approximants

Glib Yevkin [a] and Olexandr Yevkin [b]*

[a] *University Information Technology, York University, Toronto, Canada;* [b] *Research and Development, Software for Structures, Toronto, Canada*

* Contact: Alex Yevkin alexyevkin@hotmail.com

**Abstract.** The advantages and difficulties of application of Padé approximants to two-dimensional regression analysis are discussed. New formulation of residuals is suggested in the method of least squares. It leads to a system of linear equations in case of rational functions. The possibility of using Tikhonov regularization technique to avoid overfitting is demonstrated in this approach. To illustrate the efficiency of the suggested method, several practical cases from physics and reliability theory are considered.

## 1. Introduction

*Padé approximant* of order [*n*/*m*] is the rational function

$$R(x) = \frac{\sum_{i=0}^{n} \alpha_i x^i}{1+\sum_{j=1}^{m} \beta_j x^j} = \frac{\alpha_0+\alpha_1 x+\alpha_2 x^2+\cdots+\alpha_n x^n}{1+\beta_1 x+\beta_2 x^2+\cdots+\beta_m x^m} \qquad (1)$$

In particular case when *m=0,* we have a polynomial.

*Padé approximant*s are used in many applications in science and engineering (Andrianov and Shatrov 2020; Badikov et al. 1984; Baker and Graves-Morris 1996; Bellman 1964; Brezinski and Redivo-Zaglia 2020; Van Dyke 1974, 1975; Vinogradov et al. 1987; Wuytack 1979). They are widely applied in numerical analysis, theoretical physics, mechanics, critical phenomena. The most common and mathematically proved is a technique of representing a function by Padé approximant which agrees with the power series of the given function (Baker and Graves-Morris 1996). It often gives better approximation of the function than its Taylor series. It can work when the Taylor series does not converge. This technique is also used for improving asymptotic expansions (Andrianov and Shatrov 2020; Bellman 1964; Van Dyke 1974, 1975). Padé approximants appeared in Laplace transform inversion.

We will be focused on application of rational functions in regression analysis (Badikov et al. 1984; Bishop 2006; Chen and Chen 2021; Mendenhall and Sincich 1992; Vinogradov et al. 1987). Regression is an important tool of quality engineering (Candia 2019; Crocker 1990). It allows to predict system response and to analyse information in order to support quality control tasks.

Let us have *M* points in a given interval with coordinates $(x_k, F_k)$, *k=1, 2...M*. The main goal of the regression is to define approximation function $\hat{f}(x)$ at the given interval. The most popular goodness of fit criterion is the sum of squared residuals

$$S = \sum_{k=1}^{M}(\hat{f}(x_k) - F_k)^2 \qquad (2)$$

The aim is to define parameters of approximation function minimizing *S*. In the case of rational approximants, we have to solve the problem

$$\{\alpha, \beta, n, m\} = arg\left(min(S(\alpha, \beta, n, m))\right) \qquad (3)$$

For the polynomial function this leads to a system of linear equations with respect to parameters $\alpha_i$

$$\frac{\partial S}{\partial \alpha_i} = 0$$

which can be easily solved.

Rational functions are typically smoother and less oscillatory than polynomial models, however the problem of overfitting is not resolved for rational functions, as it is done in case of polynomials by introducing Tikhonov type regularization method (Bauer and Lykas 2011; Mark and Gockenbach 2016; Tikhonov 1963). In this research we suggest a new form for residuals whose minimization in case of rational function leads to linear equations and allows to apply Tikhonov regularization method to avoid unexpected oscillation of the approximant.

Still having a moderately simple form, rational functions are more general and can extend possibilities of approximation. They can take on a wide range of shapes and can be tailored to model asymptotic behavior of a function outside of given domain, therefore they have better extrapolation properties. However, solving regression problem in the form (1.3), we have to deal with significantly more complicated nonlinear equations.

There is one exceptional case when we have a system of linear equations for calculation of rational function parameters. If the function is passing through some points, then we have

$$R(x_k) = F_k, k = 1, 2 \ldots L \qquad (4)$$

where *L* is the number of points. If *n+m+1=L*, this system of linear equations can be used for calculation of *n+m+1* coefficients $\alpha_i$ and $\beta_j$. Set of *L reference points* can be selected as any subset of all given *M* points. Considering different subsets and changing *n* and *m* under condition *n+m+1=L* one can define the function minimizing the value of sum *S* as the best approximant. This methodology was applied by Badikov et al. (1984) and by Vinogradov et al. (1987) for regression analysis of experimental data in physics.

Without any doubt this approach can be used for interpolation of results obtained numerically with good accuracy. Let us have *M=21* points generated by function $sin(2\pi x)$. They are shown in Figure 1 by black signs. They divide interval [0, 1] into 20 subintervals. If we select $L=11$ refence points which are shown in Figure 1 by circles and then changing values of $n$ and $m$, we obtain the following function minimizing sum of squared residuals $S$ ($n=8$ and $m=2$)

$$R(x) = \frac{6.285x-3.396x^2-37.28x^3+17.36x^4+76.98x^5-83.92x^6+23.98x^7}{1-0.5305x+0.5305x^2} \text{ with } \alpha_0 = \alpha_8 = 0 \quad (5)$$

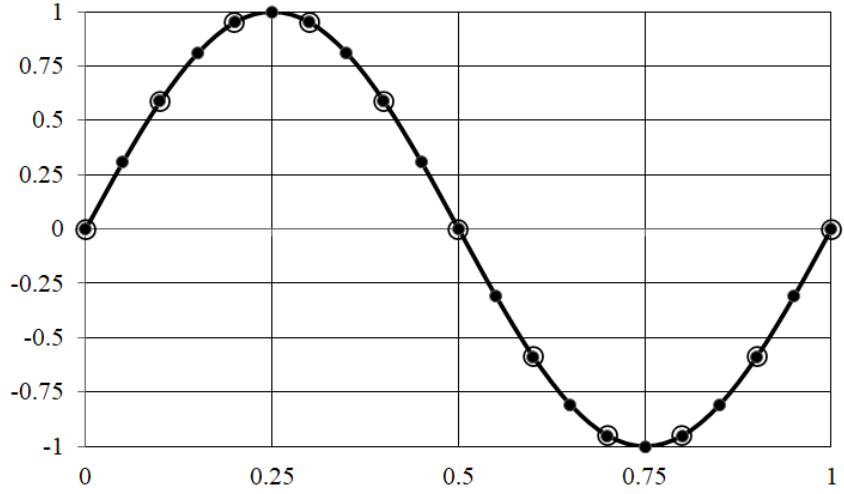

**Figure 1.** Interpolation of sinusoidal function by Padé approximant passing through reference points which are shown by circles.

As a measure of deviation of our solution from given points we use the root mean square error

$$D = \sqrt{S/M} \qquad (6)$$

For data approximation by rational function (5) we have $D = 6.324E - 6$. In case of polynomial function $n=10$ and $m=0$ we obtain

$$P(x) = 6.28x + 0.13x^2 - 42.9x^3 + 9.93x^4 + 42.5x^5 + 98.4x^6 - 235x^7 + 155x^8 - 34.5x^9 \quad (7)$$

with $\alpha_0 = \alpha_{10} = 0$ and $D = 1.247E - 5$.

Both polynomial and rational functions yield excellent interpolation of sinusoidal function, however the latter is better according to calculated values of error D. In addition, Taylor expansion of rational function (5)

$$R(x) \approx 6.285x - 0.06186x^2 - 40.64x^3$$

matches better than (7) with Taylor series of sinusoidal function

$$sin(2\pi x) \approx 6.283x - 41.34x^3$$

The described method was also applied for analytical approximation of experimental data to model different processes in physics (Badikov et al.1984; Vinogradov et al. 1987). The input data had certain error, therefore different sets of reference points were tried with different combination of numbers $n$ and $m$ while minimizing the sum of squared residuals $S$. An efficient iteration process of minimization was suggested by Badikov et al.(1984) and Vinogradov et al. (1987). However, this method requires many data points $M$ and it is limited by the level of input error which should be quite moderate.

In the next Section 2 we suggest a new way of selecting refence points which less depends on input error, but still needs a significant number of input points. In the Section 3 we suggest a new form of sum of squared residuals which leads to a system of linear equations for calculation of rational function parameters. We consider several examples to illustrate the efficiency of the method. In the Section 4 it is shown that Tikhonov regularization technique can be applied to avoid possible effect of overfitting. This is also illustrated by examples.

## 2. Reference points for rational approximants

We suggest a new way to select refence points for data approximation by rational functions. Let us have enough scattered points at a given interval. We can group them by several subsets and calculate mean values of their coordinates (of each subset) and take accordingly obtained points as reference ones for data approximation.

In our example, which is similar to example in book by Bishop (2006), we generated data randomly for further regression analysis considering underlying sinusoidal function in the form

$$F(x) = sin(2\pi x)(1 + N(x)) \qquad (8)$$

where $N(x)$ normally distributed probability function with mean $\mu$ and standard deviation $\sigma$. We put $\mu = 0$ and $\sigma = 0.1$ in this example. Obtained $M=47$ data points are shown in Figure 2 by black signs. The underlying sinusoidal function $f(x) = sin(2\pi x)$ is presented by red solid line. It can be thought of as exact solution of the regression problem. One can see that deviation of input data from underlying function is significant. It can be estimated as root mean square error

$$D_0 = \sqrt{S_0/M} \text{ with } S_0 = \sum_{k=1}^{M}(f(x_k) - F_k)^2 \qquad (9)$$

In this example we know the exact solution and therefore we can also estimate deviation of our approximation $\hat{f}(x)$ from exact function $f(x)$ calculating error

$$D_1 = \sqrt{S_1/M} \text{ with } S_1 = \sum_{k=1}^{M}(f(x_k) - \hat{f}(x_k))^2 \qquad (10)$$

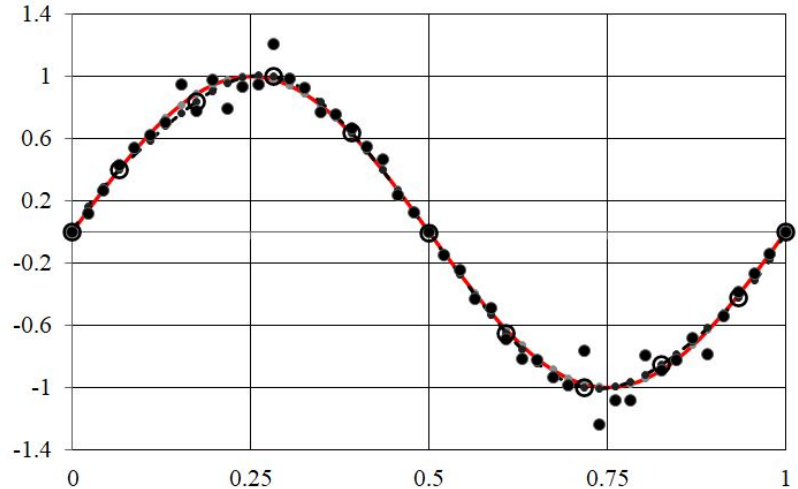

**Figure 2.** Approximation of randomly distributed data (black signs). Solid red curve is sinusoidal underlying function ('exact solution'). Dashed line represents rational approximation passing through reference points shown by circles.

First, we selected two refence points (0, 0) and (1, 0). The rest 45 points were divided by 9 subsets with 5 points each. Taking average coordinates of points from each subset, we obtained additional 9 reference points. All reference points are shown in Figure 2 by circles. They are close to underlying sinusoidal function. Then we examined rational function passing through these $L=11$ reference points solving corresponding system of $L$ linear equations (4). We considered all possible combinations of $n$ and $m$ ($n+m+1=L$) calculating sum of squared residuals $S$ (2) for all $M$ points. Minimum $S$ corresponds to the following solution $n=4$, $m=6$ with $D = 0.0840$. Deviation of input points from exact function $D_0 = 0.0838$. Deviation of the obtained approximation from exact solution $D_1 = 0.0253$ is even smaller. Corresponding dashed curve almost coincides with the underlying function. Approximation by polynomial ($n=10$, $m=0$) yields $D = 0.0859$ and $D_1 = 0.0285$.

Obviously, using rational approximants, we can obtain better regression solution compared to polynomial function. Selecting refence points (as it was suggested) allows not only to remain in scope of linear problem but also can make the solution smooth and can compensate random dispersion of data (noise). This approach was implemented in online software http://www.soft4structures.com/products.html. According to our experience, it works well if the number of points for approximation is large. In the next section we will suggest a more general approach based on a new form of sum of squared residuals. It works well for the case relatively small number of input data points.

## 3. New form of sum of squared residuals

In general case, the equations $R(x_k) = F_k$ cannot be satisfied for all $M$ given points and deviation from exact solution is minimized by minimizing sum of squared residuals (2). This leads to a system of nonlinear equations for rational function. However, we can rewrite these equations in the equivalent form

$$\sum_{i=0}^{n} \alpha_i x^i - F_k \sum_{j=1}^{m} \beta_j x^j = F_k \tag{11}$$

and then minimize the sum of squared residuals for these equations. We have

$$S_0 = \sum_{k=1}^{M} \left( \sum_{i=0}^{n} \alpha_i x^i - F_k \sum_{j=1}^{m} \beta_j x^j - F_k \right)^2 \tag{12}$$

Obviously, approximant tends to exact solution if $S_0 \to 0$. Minimization of $S_0$ by taking derivatives with respect to rational function parameters leads to a system of liner equations. Elements of matrix $A$ and vector $B$ of this system of linear equations $AX = B$ are the following if $i<n+2$ { if $j<n+2$ $A_{ij} = \sum_{k=1}^{M} x_k^{i+j-2}$ else $A_{ij} = -\sum_{k=1}^{M} F_k x_k^{i+j-n-2}$ } else {if $j<n+2$ $A_{ij} = \sum_{k=1}^{M} F_k x_k^{i+j-n-2}$ else $A_{ij} = -\sum_{k=1}^{M} F_k^2 x_k^{i+j-2n-2}$} $\tag{13}$

where $i, j = 1, 2, ..., n+m+1$ and $k=1,2, ..., M$

and

if $i<n+2$ $B_i = \sum_{k=1}^{M} F_k x_k^{i-1}$ else $B_i = \sum_{k=1}^{M} F_k^2 x_k^{i-n-1}$ $\tag{14}$

We have to mention here that in case of polynomials $S_0 = S$ therefore the form (12) coincides with traditional (2).

Transformation of initial equations before minimization of sum of squared residuals is used, for example, in reliability theory in case of estimation of parameters of Weibull lifetime probability distribution. Initial equations are the following

$$1 - e^{-(x_k/\theta)^\beta} = F_k \tag{15}$$

where $\theta$ and $\beta$ are scale and shape parameters of Weibull function subject to estimate. After taking logarithm two times we have the equivalent linear equations

$$\beta \ln x_k - \gamma = \ln \ln \frac{1}{1-F_k} \tag{16}$$

with respect to $\beta$ and $\gamma = \beta \ln \theta$. Then method of minimization of sum of squared residuals is applied to these transformed equations.

Many authors applied the Padé approximations in modeling physical processes and material properties (Badikov et al.1984; Vinogradov et al. 1987; Zang et al. 2008, 2009, 2011). These functions are used by Zang et al. (2008) for identification of air bubble volume. Padé approximations are considered as a numerical inversion method for the estimation of the quality Q factor and phase velocity in linear, viscoelastic, and isotropic media using the reconstruction of relaxation spectrum (Zang et al. 2009, 2011). Some special physical phenomena are described by rational functions (Badikov et al.1984; Vinogradov et al. 1987), for example continuous probability density function which is based on Breit–Wigner resonance formula (Breit and Wigner 1937; Brezinski and Redivo-Zaglia 2020). Therefore in many practical cases the rational functions describe the object properties and behavior according to its physical nature.

The following underlying function

$$f(x) = \frac{1}{(x+0.5)^2+0.5^2} + \frac{1+0.2x}{(x-0.5)^2+0.3^2} \qquad (17)$$

was used in the case study (Badikov et al.,1984) of regression analysis. The canonical form of this Padé function is

$$f(x) = \frac{4.941+0.5882x+12.94x^2+4.176x^3}{1-x-x^2+5.882x^4} \qquad (18)$$

To introduce "noise", this function was multiplied by factor $(1 + N(x))$ where $N(x)$ normally distributed probability function with mean $\mu = 0$ and standard deviation $\sigma = 0.05$. 21 points were generated in the interval [-1, 1] dividing it by 20 subintervals. These points are shown in Figure 3 by black signs. The underlying function $f(x)$ is presented by solid red curve.

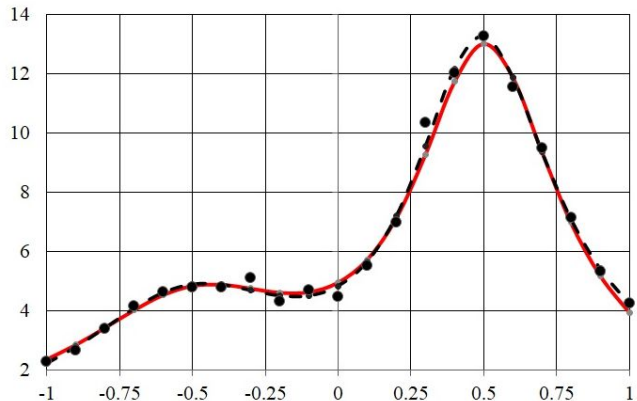

**Figure 3.** Underlying rational function (17) describing nuclear resonance represented by red solid line. 21 black points were randomly generated as an input for regression analysis. Dashed curve is data approximation by rational function (19).

We minimized the sum of squared residuals $S_0$ solving corresponding linear equations with respect to $\alpha_i$ and $\beta_j$. We repeated this procedure with different combination of $n$ and $m$ and finally selected the following solution $n=3$, $m=4$ minimizing value of $S(n,m)$

$$R(x) = \frac{4.824 + 0.5408x + 14.21x^2 + 4.379x^3}{1 - 1.023x - 1.029x^2 + 0.6705x^3 + 6.049x^4} \qquad (19)$$

We have to emphasize that we minimized $S_0$ while calculating $\alpha_i$ and $\beta_j$ for fixed $n$ and $m$. At the second part of calculation, we minimized $S(n,m)$ with respect to $n$ and $m$.

Formula (19) by its form is close to the original underlying function (18). The solution (19) is shown by dashed line in Figure 3. Corresponding error is $D = 0.2483$. Deviation of the approximation from the exact solution $D_1 = 0.1632$. The input data error $D_0 = 0.3231$.

After increasing standard deviation $\sigma = 0.1$ in this example, input error increased two times $D_0 = 0.6432$ while approximation error $D_1 = 0.289$ increased not as much for the obtained solution. We obtained $n=3$, $m=4$, $D = 0.5481$ in this case minimizing value of $S(n,m)$.

In both cases further increase of order of rational function $n$ and $m$ did not yield considerably better accuracy of the approximation. In the next Section we will discuss the problem of overfitting which can occur when the number of free parameters of approximant is increasing and causing undesired oscillation of the solution.

## 4. Applying regularization technique

We have to mention that form (11) is also used for nonparametric regression analysis by Ahmed et al. (2020). Regularization technic is applied for this ill-posed inverse problem. The solution of ill-posed problem can significantly depend on perturbations (noise). To avoid this undesired effect a Padé-type approximation based on truncated total least squares method was developed which is efficient for fitting data with relatively large number of data points. Another problem in regression analysis is an overfitting. The phenomenon of overfitting occurs when the approximation function is too closely aligned to a limited set of data points. For example, if we have 10 points for regression analysis, we can define 10 parameters of function passing through all given points, therefore minimum $S=0$. However, error of such approximation will be equal to initial error of input data, and it essentially depends on the input. In addition, the approximation function passing through given points can intensely oscillate which is undesired for a smooth function. In the case of polynomial approximation this oscillation is caused by growing values of polynomial coefficients. Tikhonov (1963) suggested to restrict these coefficients introducing terms of squared coefficients to sum of squared residuals. In our method it is

$$S_* = S_0 + \lambda \sum_{i=0}^{n} \alpha_i^2 + \lambda_1 \sum_{j=1}^{m} \beta_j^2 \qquad (20)$$

Technique of selecting appropriate value of $\lambda$ is well studied (Bauer and Lykas 2011; Bishop 2006; Mark and Gockenbach 2016) for the case of polynomial function ($\beta_j = 0$). Second parameter $\lambda_1$ can regulate coefficients of polynomial in the denominator of rational function. The denominator cannot be equal or very close to zero to avoid undesired discontinuity or oscillation of the function. Note that we still have linear regression problem with respect to coefficients of rational function. Technically, diagonal matrix should be added to matrix $A$ (13) in this method.

We considered this approach applying it to underlying Weibull probability function

$$W(x) = 1 - e^{-(x/\theta)^\beta} \tag{21}$$

which is the most popular in reliability theory. Using Monte Carlo method, we simulated times of failures of $M$ components

$$x_k = \theta\bigl(-ln(1 - P(k))\bigr)^{1/\beta}, k = 1, 2, \ldots, M \tag{22}$$

where $P(k)$ are generated random numbers in the range interval [0, 1]. Failure times $x_k$ should be sorted in ascending order. Then corresponding values of probability can be calculated using median rank formula (O'Connor and Kleyner 2012)

$$F_k = (k + 1 - a)/(M + 1 - 0.5a) \tag{23}$$

where $a = 0, 0.3, 0.5$ for different approximations of median ranks. We put $a = 0.3$ and $\theta = 1, \beta = 2$ in this example. We generated 10 points presented in Figure 4 by black signs. Their coordinates are also provided in Table 1.

**Table 1.** Coordinates of randomly generated points with Weibull underlying function

| $x_k$ | 0.3268 | 0.3426 | 0.5347 | 0.6611 | 0.7530 | 0.7933 | 1.0141 | 1.0352 | 1.2152 | 1.6970 |
|---|---|---|---|---|---|---|---|---|---|---|
| $F_k$ | 0.0673 | 0.1635 | 0.2596 | 0.3558 | 0.4519 | 0.5481 | 0.6442 | 0.7404 | 0.8365 | 0.9327 |

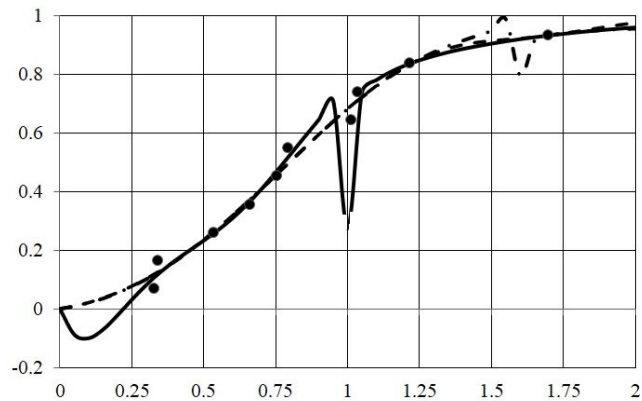

**Figure 4.** Randomly generated points were fitted by rational function using regularization technique ($\lambda = 0.0025$, dashed curve). Oscillated solid line is the result of approximation before regularization ($\lambda = 0$). Dot-dashed curve corresponds to $\lambda = 0.002$.

For approximation we tried rational function in the following form

$$R(x) = \frac{\alpha_1 x + \alpha_2 x^2 + \cdots + \alpha_n x^n + \alpha_l x^l}{1 + \beta_1 x + \cdots + \beta_m x^m + \alpha_l x^l} \text{ with } l > n, m \qquad (24)$$

satisfying the obvious conditions for probability function: $R(0) = 0$ and $R(x) \to 1$ if $x \to \infty$. Considering approximants of different order with $\lambda = 0$, we obtained optimal solution for $m = 0$, $n = 6, l = 12$

$$R(x) = \frac{-2.79x + 24.1x^2 - 72.2x^3 + 107x^4 - 76.9x^5 + 22x^6 - 1.03x^{12}}{1 - 1.03x^{12}} \text{ with } D = 0.03195 \qquad (25)$$

minimizing $S(n, m, l)$. It is shown in Figure 4 by solid line. Even though the curve is aligned very closely to input points, it is oscillating and even has vertical asymptote, hence we used regularization in the form (20) to make the solution smooth and monotonically increasing. For the case of $\lambda = 0.002$ we obtained the solution ($D = 0.032$) which is shown in Figure 4 by dot-dashed line. It is still oscillating. To select proper value of parameter $\lambda$ we introduced a measure of oscillation by calculating derivative of the approximant $R'(x)$ at $N = 40$ points inside the interval $[0, 2]$ and then evaluated the mean root squares

$$D_{der} = \sqrt{\sum_{i=1}^{N}(R'(x_i))^2 / N}, i = 1, 2, \ldots N \qquad (26)$$

The graph of function $D_{der}(\lambda)$ is plotted in Figure 5 by dashed line. The function is almost constant if $\lambda \geq 0.0025$ and it is close to the value $D_{der} = 0.5595$ corresponding to exact Weibull function. But it is increasing sharply if $\lambda \to 0$ reflecting the fact of significant oscillation of the rational function. Error $D(\lambda)$ (solid line) is slightly increasing with parameter $\lambda$. One can easily conclude that optimal value of parameter $\lambda = 0.0025$ with $D_{der} = 0.5804$ and $D = 0.03213$. Corresponding approximant is defined as

$$R(x) = \frac{0.152x + 0.579x^2 + 0.234x^3 - 0.136x^4 - 0.241x^5 + 0.0953x^6 + 0.00307x^{12}}{1 + 0.00307x^{12}} \qquad (27)$$

and plotted in Figure 4 by dashed line. It is smooth and monotonically increasing. Its denominator is always positive, and it has fewer values of coefficients compared to function (25). Solving the regression problem in this example, we put $\lambda_1 = 0$ because the result did not depend much on this parameter: the denominator in (27) is always positive and coefficient of its polynomial is small.

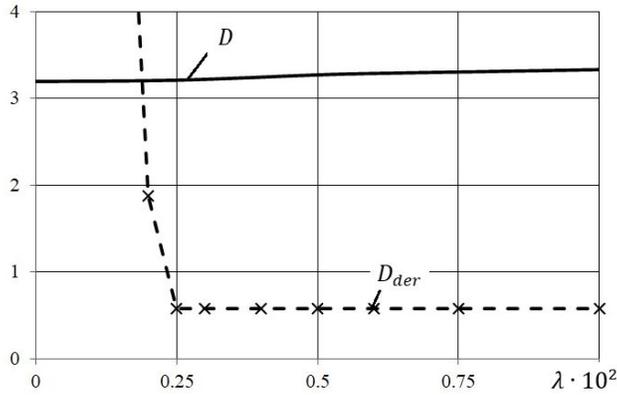

**Figure 5.** Dependence of error $D$ and oscillation factor $D_{der}$ from parameter $\lambda$ whose optimal value is $\lambda = 0.0025$.

In Figure 6 the obtained approximant is compared with the exact Weibull function (solid red curve) and with the solution obtained by maximum likelihood estimator ($\theta = 0.9488, \beta = 2.252$) which is common in reliability theory (O'Connor and Kleyner, 2012). The most important feature in reliability analysis is mean time to failure. Its exact value in our case is 0.8862, rational approximation yields 0.8327, maximum likelihood estimator - 0.8404. Both approximations yield good result despite the limited number of input data points. In case of maximum likelihood criterion, we have good approximation of input data having only two free parameters ($\theta, \beta$) because the approximation and underlying functions have the same shape of Weibull lifetime distribution.

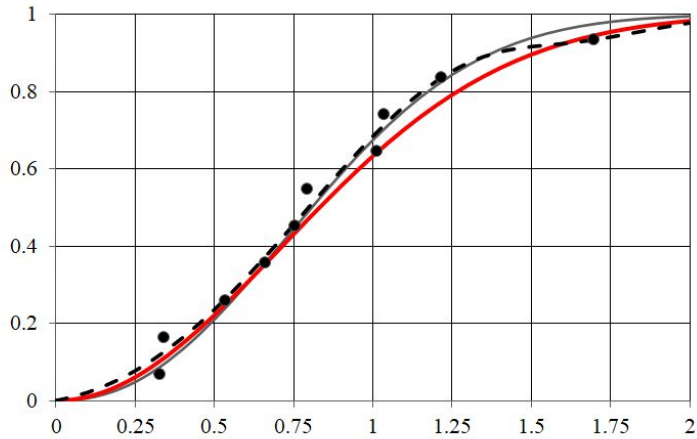

**Figure 6.** Approximation of input data (black signs) by rational function (dashed line) is compared with exact solution (thick red curve) and result obtained by maximum likelihood estimator (thin solid black curve).

In the next example we generated input data randomly considering function in the form

$$F(x) = \sqrt{x}e^{-x}(1 + N(x)) \qquad (28)$$

where $N(x)$ is a normally distributed probability function with mean $\mu$ and standard deviation $\sigma$. We put $\mu = 0$ and $\sigma = 0.1$ in this example. Obtained $M=11$ data points are shown in Figure 7 by black signs. The underlying function $f(x) = \sqrt{x}e^{-x}$ is presented by red solid line. Solving regression problem, we made the substitution $x \to x^q$ where new introduced parameter $q$ was calculated in the process of data fitting by minimizing sum of squared residuals (2). At the first step we tried polynomial function ($n = 3, m = 0$). The solution is

$$P(x) = 0.0001 + 1.341x - 1.292x^2 + 0.3273x^3 \qquad (29)$$

with $q = 0.65$. It is shown in Figure 7 by thin solid line. Deviation of this solution from given points (6) $D = 0.0206$, its deviation from the underlying function $f(x)$ (exact solution) is $D_1 = 0.0146$ (eq. 2.3). No oscillation was observed in the interval [0, 2] when $\lambda = 0$ and coefficients of polynomial (29) were not large. The interval [0, 2] was divided by $N = 100$ points and measure of oscillation was estimated by calculating derivative values of the polynomial at these 100 points inside of interval [0, 2] and then the mean root squares (26) was evaluated $D_{der} = 0.5088$. The value corresponding to exact solution $D_{der} = 0.5268$.

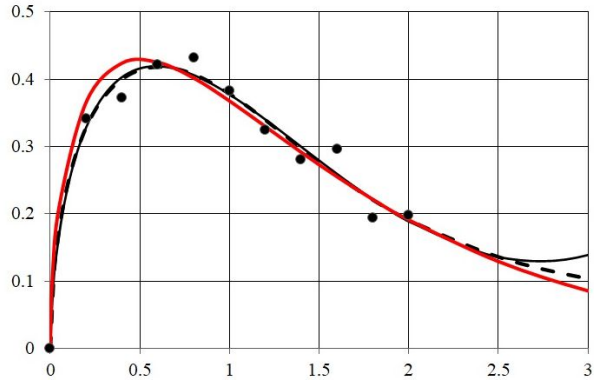

**Figure 7.** Data fitting with regularization. Red solid line is underlying function, thin curve is polynomial approximation ($m = 0, n = 3$) and dashed line is rational approximant ($m = 6, n = 3$) which yields better data extrapolation.

We have already obtained not a bad data approximation at the interval [0, 2] by polynomial and hence we can use this result for the next step of data fitting. From Figure 7 one can see that the approximation is far from exact solution, and it is even increasing when $x > 2.8$. To fix this undesired behavior, we tried rational functions with $m > n$ and obtained solution with horizontal asymptote $y = 0$. In the case when $m = 6, n = 3$ we have the rational function

$$R(x) = \frac{1.261x - 1.155x^2 + 0.3922x^3}{1 + 0.0251x + 0.0316x^2 + 0.0404x^3 + 0.0528x^4 + 0.0708x^5 + 0.097x^6} \text{ with } D = 0.02, D_1 = 0.0144 \quad (30)$$

which is shown in Figure 7 by dashed line. It yields a much better extrapolation of the underlying function. The following values $q = 0.6$ and $\lambda_1 = 0.4$ minimized the sum of squared residuals

(minimum $D$). Selecting value of regularization parameter $\lambda_1$, we also checked value of $D_{der}$ which must not differ much from value 0.5088 obtained for polynomial function. If it is significantly greater, then the solution is oscillating or even has unexpected vertical asymptotes as it really happened when we tried to put $\lambda_1 = 0$. We started the process with $\lambda_1 = 1$ and came to optimal value $\lambda_1 = 0.4$ with $D = 0.02$ and $D_{der} = 0.4997$. All coefficients of polynomial in the denominator (30) are positive in this case. We believe that technique of defining polynomial coefficients of the denominator (selecting optimal value of $\lambda_1$) as well as solving extrapolation problem is worth developing in future.

## 5. Conclusion

Suggested in Section 2 new method of refence points selection for regression analysis using rational functions is rather simple and leads to a system of linear equations. As it is illustrated by example, it can significantly compensate the effect of dispersion of data and make the approximation function smooth. However, it requires many data points to gain the benefits of the approach.

For smaller number of points, it is better to use the new definition of residuals to minimize their sum in regression analysis developed in this paper. This approach also leads to a simple system of linear equations. It allows trying Padé functions of different order and combinations of $n$ and $m$. Polynomials are included automatically in the case when $m=0$. In addition, such substitutions as $x \to x^q, \ln x, \exp(qx)$ with additional parameter $q$ can be incorporated in the regression analysis. The simplicity of calculation of Padé function coefficients allows to consider different combinations of other parameters of the model $(n, m, q, \lambda)$ minimizing sum of squared residuals $S$. It is also important that asymptotic properties of the expected solution can be rather easily incorporated in the rational approximants.

Remaining in the scope of linear problem, Tikhonov regularization technique can be used, which is very important in many practical cases. We believe that theoretical background of the suggested methods and regularization procedure in case of rational function should be developed as a next step of the research.

Considering several examples, we have suggested that at least in two cases the Padé functions are most suitable for regression analysis: when some asymptotic properties of the underlying function are known and when the object is described by Padé function according to its nature. The statistical properties of the object can be predicted by the regression model and therefore used in different applications.